\documentstyle[emulateapj,flushrt,tighten]{article} %preprint
\input{psfig.sty}

\def\BS{{\it Beppo}SAX~}

\def\fluxunits{\,{\rm ergs\,cm^{-2}\,s^{-1}}}
\def\gta{ \lower .75ex \hbox{$\sim$} \llap{\raise .27ex \hbox{$>$}} }
\def\lta{ \lower .75ex\hbox{$\sim$} \llap{\raise .27ex \hbox{$<$}} } 

\def\preprint{
}

 1
\begin{document}
\preprint

\title{Broad-band X-ray spectra of the persistent black hole candidates
 LMC X--1 and LMC X--3}

\author{
F. Haardt\altaffilmark{1}, 
M.R. Galli\altaffilmark{1}, 
A. Treves\altaffilmark{1}, 
L. Chiappetti\altaffilmark{2}, 
D. Dal Fiume\altaffilmark{3\dagger}
A. Corongiu\altaffilmark{4,2},
T. Belloni\altaffilmark{5},
F. Frontera\altaffilmark{6,3},
E. Kuulkers\altaffilmark{7,8},
L. Stella\altaffilmark{9,10}
}
\altaffiltext{1}{Dipartimento di Scienze, Universit\`a dell'Insubria, Como, 
Italy}
\altaffiltext{2}{IFC "G. Occhialini", CNR, Milano, Italy}
\altaffiltext{3}{ITESRE, CNR, Bologna, Italy. $\dagger$ {\it Prematurely deceased, 5 August 2000}}
\altaffiltext{4}{Dipartimento di Fisica, Universit\`a di Milano, Milano, Italy}
\altaffiltext{5}{Osservatorio di Brera, Merate, Italy}
\altaffiltext{6}{Dipartimento di Fisica, Universit\`a di Ferrara, Ferrara, Italy}
\altaffiltext{7}{SRON, Utrecht, The Netherlands}
\altaffiltext{8}{Astronomical Institute, Utrecht University, The Netherlands}
\altaffiltext{9}{Osservatorio Astronomico di Roma, Monteporzio Catone, Italy}
\altaffiltext{10}{Affiliated to ICRA}
%\slugcomment{submitted to ApJ 2000 July 17}
\received{---------------}
\accepted{---------------}

\begin{abstract}
We report on observations of the two persistent black hole 
candidates LMC X--3 and LMCX--1 performed with \BS in October 1997.
The flux of 
LMC X--1 was possibly measured up to 60 keV, but there is a possible confusion with
PSR 0540-69. 
Fits with an absorbed multicolor disk black body are not satisfactory, 
while the superposition of this model with a power law is acceptable. 
The sources showed little variations during the observations. 
However in LMC X--1 some X--ray color dependence on intensity is apparent, 
indicating a hardening of the spectrum in the second half of the observation. 
The inner disk radius and temperature change, featuring the same (anti)correlation 
found in {\it RXTE} data (Wilms et al. 2000). QPOs were searched for. 
In LMC X--3 none was detected; in LMCX--1 a 3 $\sigma$ upper ($\simeq 9\%$ rms) 
limit is given at 0.07 Hz, the frequency of the QPO discovered with Ginga. 
\end{abstract}
%\twocolumn

%%%%%%%%%%%%%%%%%%%%%%%%%%%%%%%%%%%%%%%%%%%%%%%%%%%
\section{Introduction}

Together with Cyg X--1, LMC X--1 and LMC X--3 are the only persistent X--ray 
binaries where the presence of a black hole is established by accurate 
measurement of the mass function.
The binary nature of LMC X--3 was discovered by Cowley et al. (1983), who 
observed an optical period of 1.7 days, and derived a mass function 
$f(M)= 2.3 M_{\odot}$. 
The likely mass of the compact object is $M_{\rm X}\simeq 9 M_{\odot}$, 
strongly indicative of a black hole. In the case of LMC X--1, 
an estimate could be obtained only after an accurate X-ray position was derived from 
ROSAT observations (Cowley et al. 1995). 
The optical period is $4.2$ days, the mass function is $f(M)=0.14 M_{\odot}$, the inferred 
mass of the compact star is $M_{\rm X}\simeq 6 M_{\odot}$.

Since their discovery with Uhuru (Leong et al. 1971), the two sources have 
been studied practically with all the X--ray missions. 
Both sources are usually found in the so--called soft--high state. 
Wilms et al. (2000, hereinafter W00) described the observation of sporadic
episodes of low/hard state in LMC X-3, and recently Boyd \& Smale (2000) 
and Homan et al. (2000) reported clear signs that the source entered a low/hard state.
In comparison, the other two persistent black-hole candidates (BHCs), Cyg X--1 
and GX~339--4, are found most of time in their hard--low state. 
We note, however, that such dichotomy of states describes only roughly the complexity of the spectral 
behaviour of BHCs, even of the persistent ones. 
As an example, the so called "intermediate state" is now known to be quite common in Cygnus X--1. 
For a review of BHC states see Tanaka \& Lewin (1995) and van der Klis (1995).

The X-ray spectrum of LMC~X--3 has been studied in detail with EXOSAT and 
GINGA (Treves et al. 1988, 1990; Ebisawa et al. 1993). 
It is consistent with the ``standard'' model, i.e., the 
superposition of a ``multicolor disk blackbody" 
(DBB) and power law high energy tail (PL), which appeared sporadically in 
the GINGA exposures. 
Such spectral shape is confirmed by recent {\it RXTE}
observations (Nowak et al. 2000, hereinafter N00; W00). 
As in most high--state BHCs, the X--ray flux of LMC X--3 is fairly stable on short 
timescales ($\lta 1$ ks), but it is subject to irregular or quasi periodic 
variability by up to a factor $\sim 4$ on 100--200 day time scales, 
as apparent in particular from the ASM--{\it RXTE} monitoring. 
Associated spectral transitions have been detected
on such long flux variations (W00). 

Also the spectrum of LMC X--1 was studied in detail with GINGA (Ebisawa et al. 
1993). 
Similar to the case of LMC X--3, it is fit by a DBB+PL, 
which agrees with recent {\it RXTE} observations by Schmidtke et al. (1999). 
The X-ray flux of the source appears to be secularly more stable than 
that of  LMC X--3, while more variable on shorter timescales (N00, W00). 
Ebisawa, Mitsuda \& Inoue (1989) reported a QPO at 0.075 Hz. The QPO was not detected
in {\it RXTE} observations (Schmidtke et al. 1999, N00).

Here we report on $\sim 40$ ks observations of LMC X--1 and LMC X--3 
obtained with \BS in October 1997. 
The main advantage of \BS compared to other missions is the broad energy 
band covered by 
its instrumentation (0.1--300 keV). Preliminary results were presented by 
Treves et al. (2000), and by Dal Fiume et al. (2000). 
The paper is structured as follows: in Section 2 we describe the data reduction, 
in Section 3 and 4 we present the data analysis for the two objects, and
in Section 5 we summarize and discuss our results.

All errors are 90\% confidence for one parameter, unless otherwise 
indicated.

%%%%%%%%%%%%%%%%%%%%%%%%%%%%%%%%%%%%%%%%%%%%%%%%%%
\section{The data}

The \BS scientific payload includes four coaligned Narrow Field
Instruments covering the nominal energy range 0.1--300 keV, namely the Low
Energy Concentrator Spectrometer (LECS, 0.1--10 keV), the 3 units of the Medium
Energy Concentrator Spectrometer (MECS, 1.3--10 keV), the High Pressure Gas
Scintillation Proportional Counter (HPGSPC, 4--120 keV) and the Phoswich
Detector System (PDS, 12--300 keV). For further details on the \BS mission
and instruments see Boella et al. (1997) and references therein.

As part of our program on persistent BHCs, \BS
observed LMC X--1 and LMC X-3 on 1997 October 5-6 (from 20:55 to 12:50 UT),
and 1997 October 11 (from 10:15 to 03:48 UT), respectively.
The data reduction procedure for the LECS, MECS, PDS and HPGSPC was 
based on the XAS software (Chiappetti \& Dal Fiume 1997; Chiappetti et al. 1999). 

The LECS and MECS are imaging instruments, thus requiring spatial selection of 
source photons. 
The extraction radius for the LECS spectra is 8 arcmin for the MECS spectra 
is 8.36 arcmin. At the time of the observations, only two MECS units were 
active. 

The exposure times and count rates are reported in Table 1.  
We note that for LMC X--3 no flux is detected in the PDS 
(the $3 \sigma$ upper limit is $\simeq 11$ cts/s in the 15--60 keV band).
LMC X--1 was detected in the PDS, but see section
5 for a discussion of possible background contamination and source
confusion. An estimate of the overall count rate corresponding to our best
background subtraction is $0.25\pm 0.03$ cts/s in the 15--60 keV range.

Because of possible source confusion in the LMC field (see below) and because 
of the low detected count rate, we did not include the HPGSPC in our spectral 
analysis.

%%%%%%%%%%%%%%%%%%%%%%%%%%%%%%%%%%%%%%%%%%%%%%%%%%% 
\section{Data analysis: LMC X--3}

\subsection{Spectral analysis}

LECS and MECS spectra were restricted to the 0.2-4 keV, and 1.8-10 keV energy 
ranges, respectively.
%\footnote{See the URL at www.asi.sax.it}.
A normalization factor was used as a free parameter in the fitting procedure, 
in order to allow for uncertainties in the cross--calibration of the LECS  
and MECS. 
The best--fit value of such parameter ($\simeq 0.75$) is in agreement 
with the nominal value indicated by the \BS SDC. 
Moreover, we include a systematic error of 2\% in the MECS data. 

\begin{figure}
\psfig{file=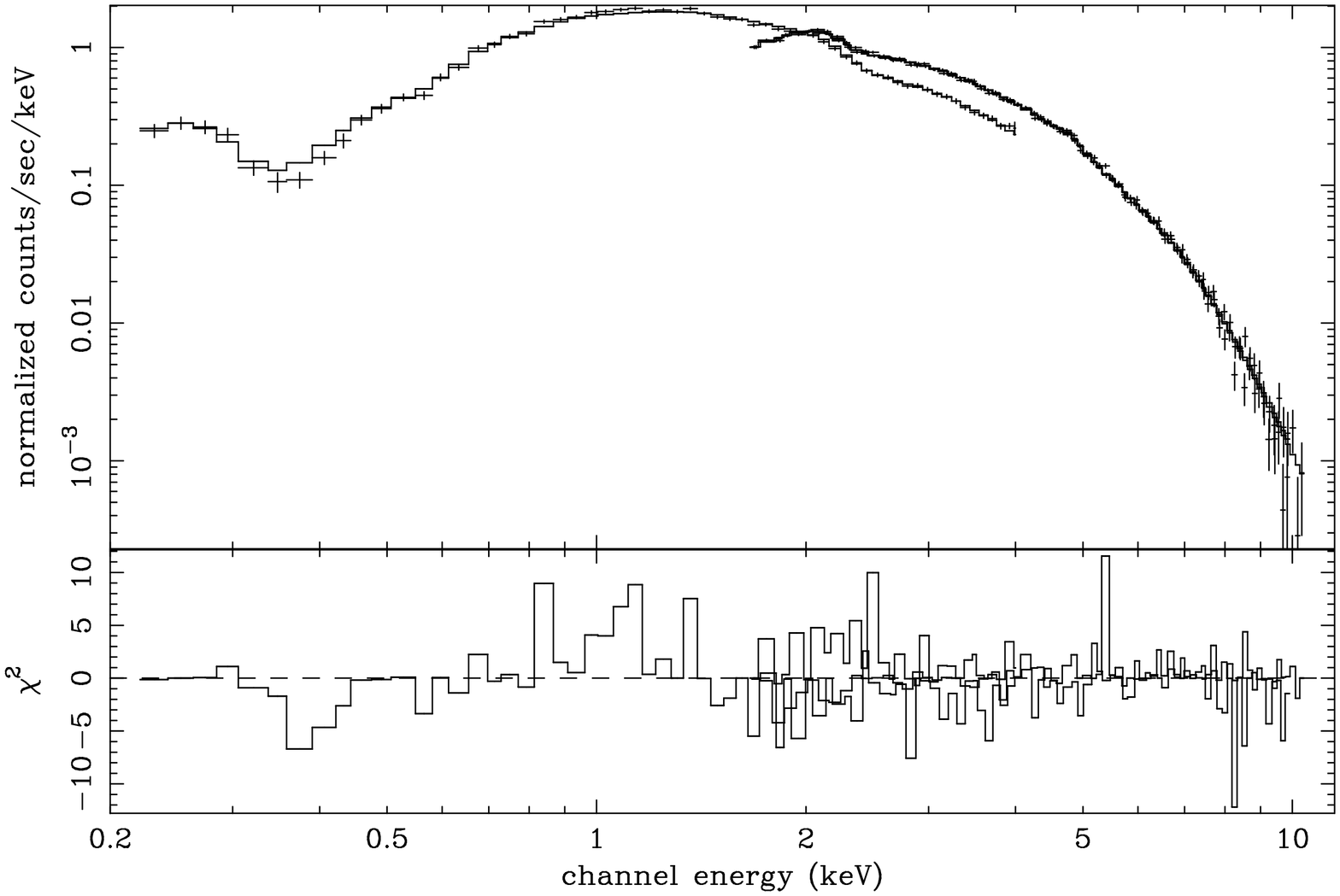,angle=270,width=1.0\textwidth}
\caption{LMC X--3:
count spectrum and contribution to $\chi^2$ when the overall data are fitted
with an absorbed DBB+PL model.}
\label{x3ff}
\end{figure}

A spectral fit with a simple absorbed DBB is not satisfactory 
($\chi^2/dof=283/216$, see Table 2). 
The model understimates the LECS data between 1 and 2 keV. 
We have therefore introduced an additional spectral component, represented by 
a PL, the inclusion of an extra component improves the fit (see Table 3). 
The spectra and the residuals are plotted in Figure \ref{x3ff}. 
The extrapolation of the power--law above 10 keV is consistent with the PDS upper limit. 

The spectral index ($\Gamma \simeq 2.7$) is close to the measurements of 
other missions (Treves et al. 1990, W00). N00 reported a slightly 
steeper power--law index ($\Gamma\simeq 3$, {\it RXTE} data), with a normalization which
is, however, a factor $\sim 20-30$ larger.   

The best fit values of the hydrogen column density 
are consistent with the mean Galactic value 
($N_{\rm H} \lta 6 \times 10^{20}$ cm$^{-2}$, Dickey \& Lockman 1990).
This result agrees with previous determinations (Treves {\it et al.} 1986, 
Treves {\it et al.} 1988a).

For what concerns the DBB component, the best fit values of the temperature 
at the inner edge of the disk (Table 2, and Table 3) are similar to 
those determined from Ginga data (Treves et al. 1990), while 
the values are lower than those derived from \BS PV (Performance Verification) 
data (Siddiqui et al. 1998), and RXTE data (W00, N00). 
Comparing our 2--10 keV flux ($F_{\rm X} = 2.7 \times 10^{-10} \fluxunits$)  
with the value reported by Siddiqui et al (1998) ($F_{\rm X} = 4.8 \times 10^{-
10} \fluxunits$, and $F_{\rm X} = 9.2 \times 10^{-10} \fluxunits$ in October 
and November 1996, respectively) shows that $T_{\rm in}$ is positively 
correlated with the luminosity. 
Our value of $R_{\rm in}$ determined in the case of DBB+PL is 
consistent with the value derived from the same model fit to Ginga data 
(Treves et al. 1990) and \BS PV data (Siddiqui et al. 1998), but it is 
smaller than that derived by N00 for RXTE data.

We did not find any evidence of emission features, such as the 6.4 keV Fe $K_{\alpha}$ line, 
in contrast to 1996 {\it RXTE} observations reported by N00. The upper limit on the 
line EW ($\simeq 300$ eV at 90\% level) is however consistent with the N00 results.

\subsection{Timing analysis}

The MECS data were subdivided in 17 separate intervals of $\sim 2000$ s 
duration, and the power spectral densities were computed for each interval, 
and then averaged together. 
No evidence for variability in excess of the counting statistics was found. 
We can then put an upper limit of $\simeq 1$ \% to the rms of variability, in the frequency 
range $10^{-3}-0.2$ Hz. 
This is in agreement with previous measurements (Treves et al. 1988, 1990; 
Schmidtke et al. 1999; N00).

%%%%%%%%%%%%%%%%%%%%%%%%%%%%%%%%%%%%%%%%%%%%%%%%%%%%%%%%%%%%%%%%%%%%%%%

\section{Data analysis: LMC X--1}

\subsection{Spectral analysis}      

The spectral analysis of LECS and MECS 
data of LMC X--1 was performed in the same way as described above. 
The source was detected in the PDS up to $\sim 60$ keV, with a significance 
$\gta 2.5\sigma$. However one should note that the 59 ms Crab
like radio pulsar PSR 0540-69 is only 25$'$ away from LMC X--1, 
and hence within the MECS field of view 
but not in the LECS, due to its narrower field of view. 
The pulsar can contribute to the flux in the PDS. 
This problem will be treated in detail in section 5. 

For the moment we consider the LECS and MECS data only, integrated over the entire 
duration of the observation. 
This spectrum is referred to as "total" in the tables. 
  
\begin{figure}
\psfig{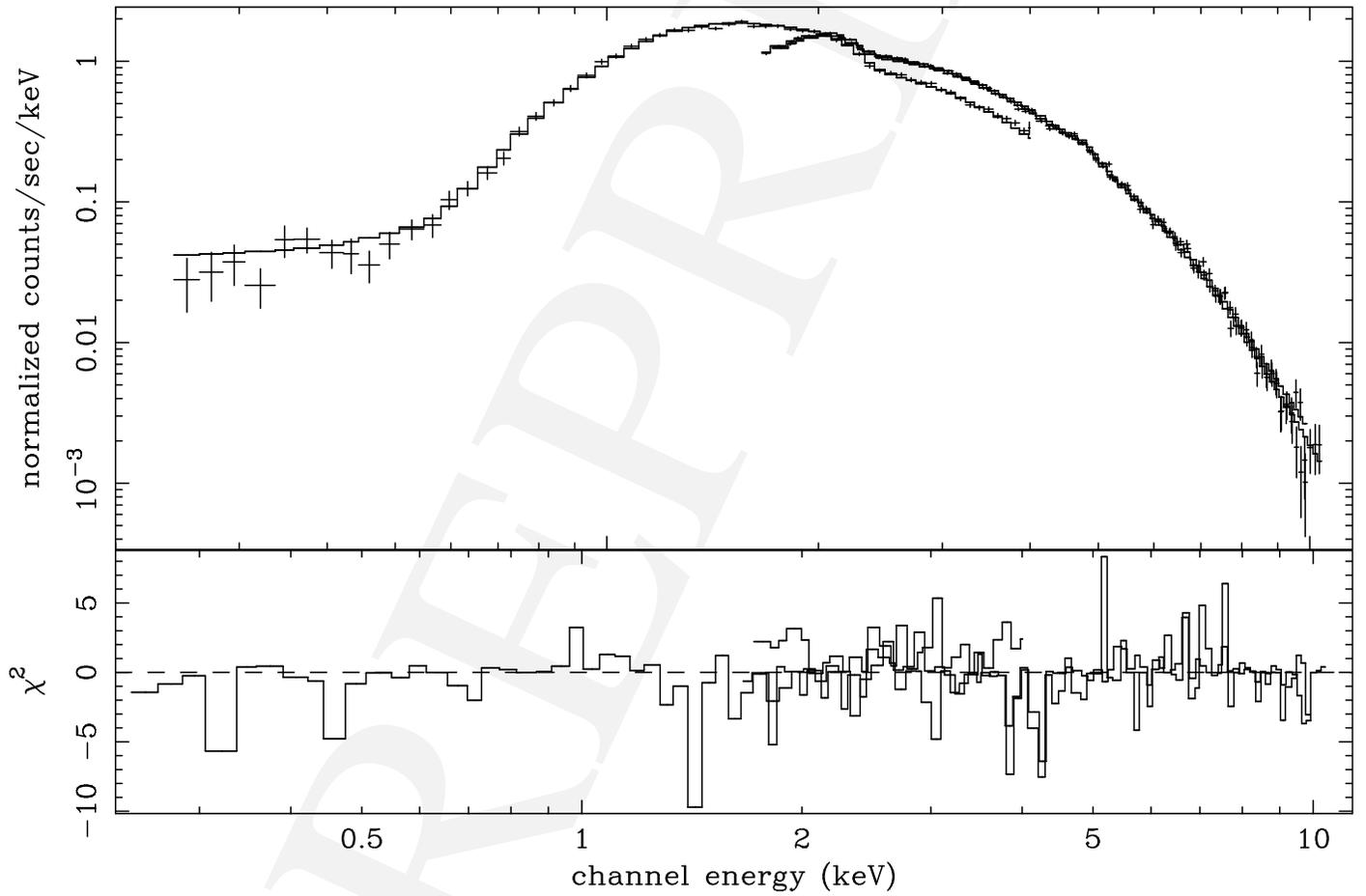}
\caption{LMC X--1:
count spectrum and contribution to $\chi^2$ when the overall data are fitted
with an absorbed DBB+PL model.} 
\label{x1po}
\end{figure}

As for LMC~X-3, a single absorbed DBB does not give an acceptable fit 
($\chi^2/dof = 708/216$, see Table 2), and a second component
is necessary in order to obtain a satisfactory fit. 
The inclusion of a PL gives $\chi^2/dof=212/193$. 
Results of the fits performed with the DBB+PL model are presented in Table 3. 
The spectrum is shown in Figure \ref{x1po}. 
Our data do not require the presence of the iron emission line at $\simeq 6.4$ keV, 
as was instead suggested by other observations (Ebisawa et al. 1989; 
Schlegel et al. 1994; N00). However our upper limit on the EW ($\simeq 250$ eV at 90\% 
level) is consistent with earlier positive detections.   

For consistency, we require that, as in the case of LMC X--3, the joint fit 
of LECS and MECS data to the DBB+PL model extrapolates in the range of 
sensitivity of the PDS without predicting to high of a flux. 
The model--extrapolated count rate in the PDS is only $(2.55\pm 1.75)\times10^{-2}$ cts/s, 
while the measured count rate is $0.25 \pm 0.03$ cts/s, in the same range (Table 1). 
This suggests that some additional contribution, possibly the already mentioned pulsar, 
must be considered to account for the total signal measured in the PDS.
 
The best fit values of the inner radius and temperature   
are consistent with those derived from {\it Ginga} data (Ebisawa et al. 1989).
{\it RXTE} data shows a similar temperature, but a larger DBB normalization (N00). 
The PL slope and normalization are similar to those derived by N00 from 
{\it RXTE} data. 
 
\subsection{Spectral variability}

\begin{figure}
\center{
\psfig{file=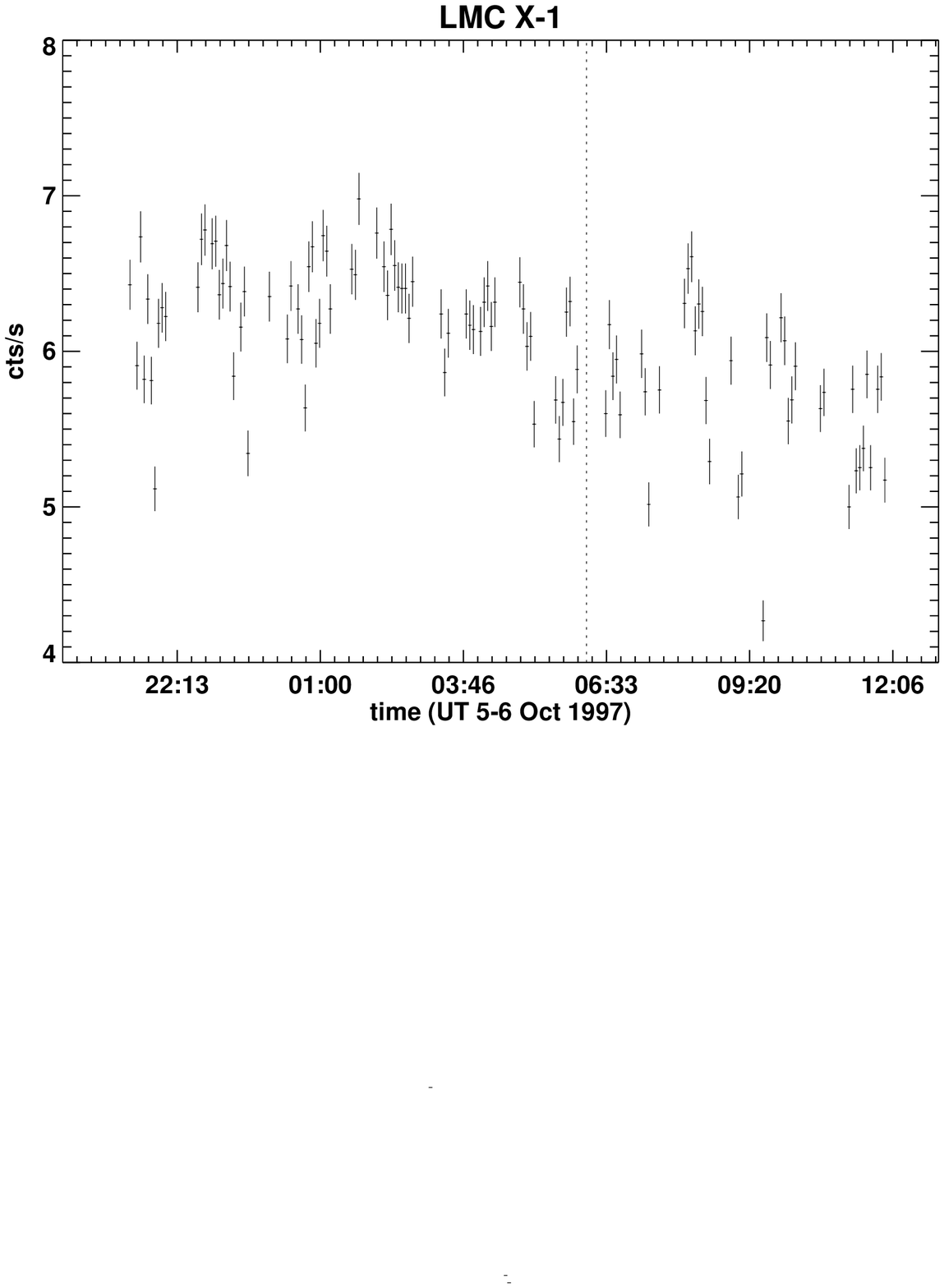,angle=0,width=1.0\textwidth,height=0.5\textheight}
\psfig{file=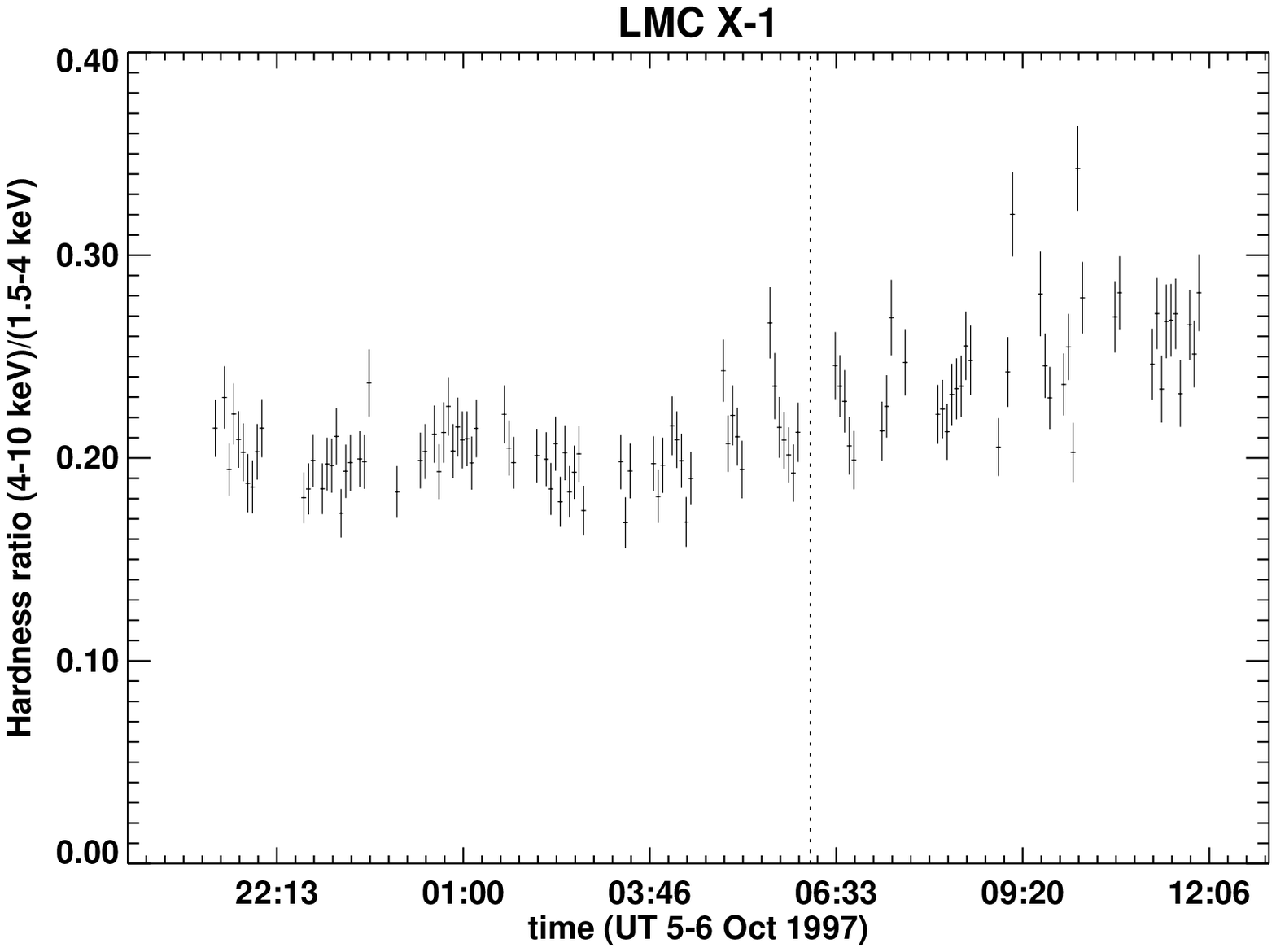,angle=0,width=1.0\textwidth,height=0.5\textheight}}
\caption{Upper panel: summed count rate of LMC X--1 in the MECS2 and MECS3 telescopes 
(1.5--10 keV energy range). The integration time is 500 s. Lower panel: 
hardness ratio versus observing time. Data left (right) to 
the vertical dashed line are Part 1 (Part 2) spectrum.} 
\label{x1lc}
\end{figure}

In Figure \ref{x1lc} (top panel) we show the lightcurve, in Figure \ref{x1lc} (bottom
panel) a color ($4-10$ keV/$1.5-4$ keV) vs. time plot. 
The last part of the observation clearly shows 
a decrease of the MECS count rate, with associated spectral evolution (hardening). 
In order to quantify such spectral evolution, we extracted two different 
spectra, collecting data from the first $\sim 30$ ks of the observation (Part 1), and  
from the remaining time (Part 2). Results are reported in Table 3. 
There are significant differences in the DBB parameters, 
with an increase of the temperature ($\simeq 9$\%) as the radius decreases ($\simeq 38$\%) 
in Part 2. 
The anticorrelation between the disk temperature and radius is similar to that observed in 
{\it RXTE} data (W00). There is also a weak indication of a hardening of the PL in Part 2. 
It should be noted that the fit to Part 2 data is quite poor in terms of $\chi^2$, so that 
it is difficult to quantify statistically such differences.    

\subsection{Timing analysis}

\begin{figure}
\psfig{file=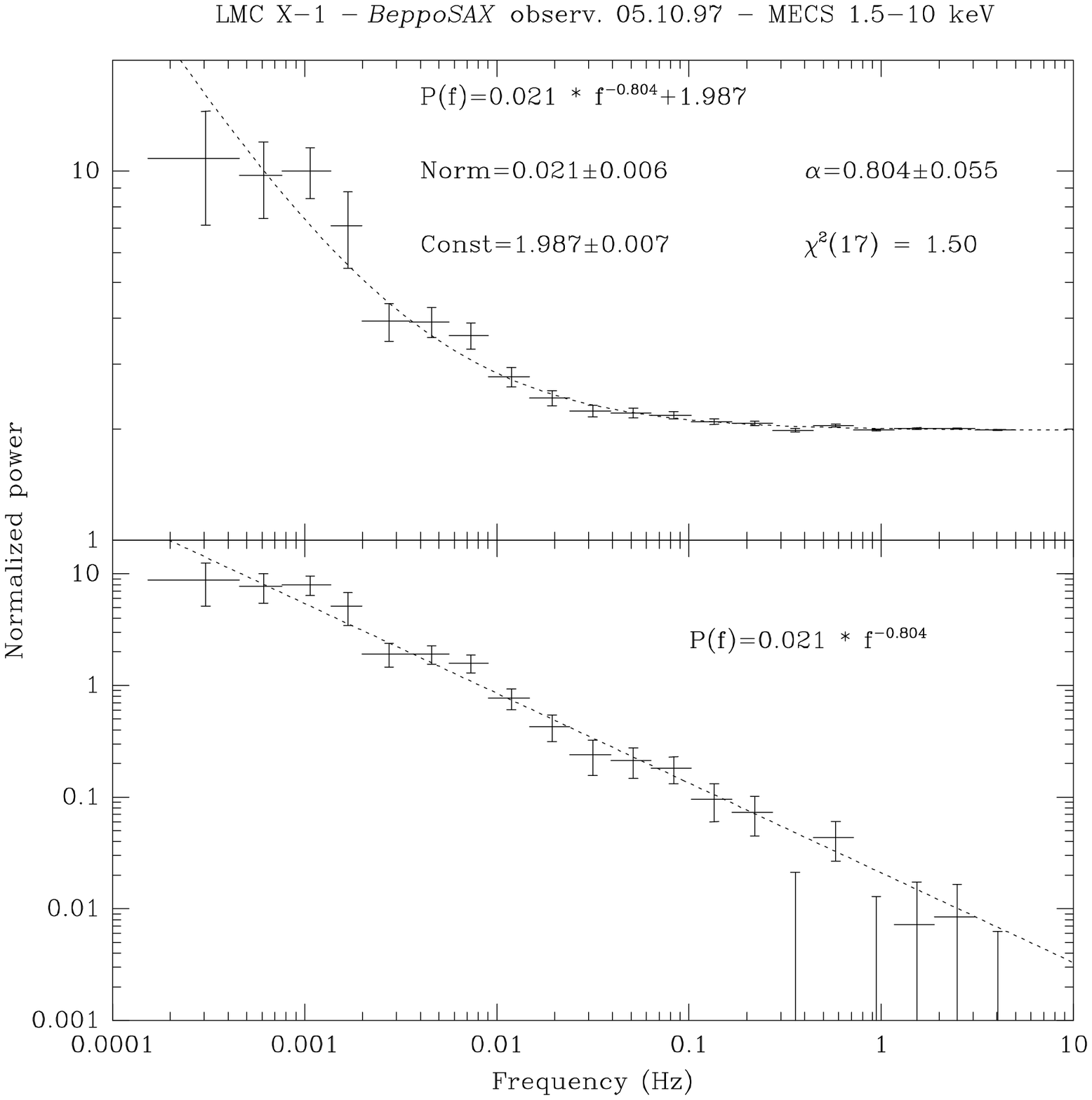,width=1.0\textwidth}
\caption{Power Spectral Density of LMC X--1 in the 1.5--10 keV band. 
A fit with a power law plus constant is also shown. Errors
are 68\% single parameter confidence level.}
\label{x1PSD}
\end{figure}

The MECS light curve of LMC X--1 shows evidence of variability (Figure \ref{x1lc}). 
We calculated PSDs from 16 uniterrupted time intervals ($\sim 3000$ s long) 
in the 1.5-10 keV, and averaged them together. 
The PSDs were normalized after Leahy et al. (1983). 
We fitted the average PSD with a power-law plus a constant to take into account the 
contribution from counting statistics. 
The power spectrum, after subtraction of the constant term, is shown in Figure 
\ref{x1PSD}.
The value of the observed slope $\alpha$ is consistent with that 
obtained in previous measurements (Ebisawa et al. 1989, Schmidtke et al. 1999, 
N00). The total fractional rms in the $3\times 10^{-4}-0.2$ Hz band is $\simeq 6\%$.
No broad line feature is evident in the PSD. We can put a 3$\sigma$ upper limit of $\simeq 9\%$ to 
the fractional rms of a QPO as the one reported by Ebisawa et al. (1989) (see 
Dal Fiume et al. 2000).

\subsection{PDS contamination from the nearby source PSR 0540--69}

Fit to LMC X--1 PDS data alone with a simple power--law yields $\Gamma=2.1^{+0.8}_{-0.5}$. 
As already noted, the source field may be contaminated by PSR 0540-69, 25' 
away from LMC X--1. 
At such distance, the PDS effective area is 68\% of that on--axis.
Unfortunately the signal from the pulsar (Mineo et al. 1999) prevents 
a pulsation analysis to asses the amount of contamination of the LMC X--1 PDS 
data. Therefore, in order to estimate the possible contribution 
of PSR 0540-69 to the PDS signal of LMC X--1, we have performed 
the analysis in two different ways, and cross--checked the results. 

The employed methods are detailed in Appendix A. Our final conclusion is that 
it seems likely that both sources contribute significantly to the PDS flux, 
but the present data are insufficient to asses their relative importance with acceptable 
significance. 
In any case, our analysis casts doubts on previous detections of a hard 
($\Gamma\simeq 2$) high energy powerlaw in LMC X--1, such as that reported by 
Ebisawa et al. (1989). 

%%%%%%%%%%%%%%%%%%%%%%%%%%%%%%%%%%%%%%%%%%%%%%%%%%%%%%%%%%%%%%%%%%%
\section{Summary and Discussion}

We have presented an analysis of \BS observations of LMC X--1 and LMC X--3. 
Here we summarize and discuss our main results.

1) The LECS and MECS spectra of both sources can be reproduced
as a superposition of a multicolor disk blackbody with a steep power law.
The parameters defining the DBB are similar to those obtained by {\it RXTE}, 
though in the case of LMC X--3, we derived significant lower temperature and normalization, 
by 20\% and 70\%, respectively. 
Power law slope values are consistent with 
the {\it RXTE} data (N00, W00), though in the case of LMC X--3 the 
normalization we obtained lies a factor $\sim 20$ below. This would imply that large 
variations of the hard component, probably indipendent on the soft component, 
may occur in this source on long time scales. 
We note that W00 reported on spectral transitions occuring in LMC X--3 on long time scales. 

2) While LMC X--3 does not show significant variability on time scales $\lta 1$ ks, 
LMC X--1 shows flux and spectral variations on such time scales. 
The trend we observe can be roughly described as  
a hardening of the spectrum as the 2--10 keV count rate decreases. Such trend is common in X--ray 
binaries. The hardening 
is due to the anticorrelation between the disk temperature and the disk radius (and a 
possible hardening of the PL component as the disk temperature increases). 
This is consistent with the idea that moving inwards the disk becomes hotter. However, it is 
not clear why the disk in LMC X--1 becomes hotter (and pushes closer to the black hole) 
while its overall luminosity decreases. 
The same behaviour is observed in {\it RXTE} data. 
Our Part 1 and Part 2 disk best fit parameters falls exactly on the $kT_{\rm DBB}-R_{\rm DBB}$ 
correlation shown by W00 (see their Figure 7). 
The \BS lower energy boundary with respect to {\it RXTE} allows, in principle, a more 
accurate constraint on such correlation. Temperature--radius contours show 
a systematic correlation between the two parameters. Therefore
we can not exclude that part of the $kT_{\rm DBB}-R_{\rm DBB}$ correlation 
is due to systematic errors in the model fit. 
W00 also report a decrease of the overall 2--20 keV flux as the disk temperature 
decreases (and the disk radius increases), mainly 
caused by a decrease (by a factor $\sim 2$) of the powerlaw flux, and argue that this fact may 
indicate the beginnings of a transition to a low/hard state. We must note that the disk 
luminosity, instead, increases by a factor $\sim 1.6$ during such transition, a behaviour 
hardly fitting in the state transition hypothesis. Our data show a different situation, since 
the powerlaw flux remains almost constant in Part 1 and 2. 

3) If the steep power law is due to Comptonization of disk photons in an active 
corona, the best fit values of the photon index, $\Gamma = 3.3$ and 
$\Gamma = 2.7$ for LMC X--1 and LMC X--3 respectively, imply 
that in these systems the Compton parameter $y\ll 1$, so that such component is 
only marginal in the overall energy budget (see also N00). 

4) We do not detect any emission line. In particular, the EW of the 6.4 keV iron line has 
an upper limit of 250 and 300 eV in LMC X--1 and LMC X--3, respectively, consistent with 
the recent {\it RXTE} data (N00).  

5) We derive only an upper limit to the presence of a QPO in LMC X--1. This result is 
consistent with the negative results of {\it BBXRT} (Schlegel et al. 1994), and {\it RXTE} 
(Schmidtke et al. 1999, N00). As N00 argue, the previously reported QPO could have 
been an artifact of misesitimation of Poissonian noise level, rather than an indication 
of a variable nature of such QPO. 

6) The pulsar PSR 0540-69 can dominate PDS counts in the LMC X--1 field, or, at least, 
give a similar contribution. 
The question arising here is whether the pulsar emission could contribute to the counts in the 
{\it RXTE} observations as well, as the angular distance from LMC X--1 is well within the 
field of view of {\it RXTE}. N00 report a steep power law ($\Gamma\simeq 3$), which is clearly 
consistent with our MECS results, but not with the pulsar emission, 
the latter featuring a much 
harder power law ($\Gamma\simeq 2$). \BS and {\it RXTE} high energy results are not, however, 
in contradiction if we assume that both power laws exist. In fact, given their relative  
normalizations (the pulsar power law, at 1 keV, lies a factor $\sim 20$ below 
the LMC X--1 one), 
it is easy to see that the pulsar and LMC X--1 emission are equal at $\simeq 20$ keV, 
the latter dominating the power law flux below. Note that the maximum energy 
{\it RXTE} detected LMC X--1 is indeed $\simeq 20$ keV.

\acknowledgments 
FH, AT and TB thank the Italian MURST and the European Union for financial support under the 
grants COFIN98-02-154, and CHRX-CT93-0329 (TMR "Accretion onto compact objects"), 
respectively. 

\appendix
\section{LMC X--1 PDS contamination from PSR 0540-69}

In order to asses the possible contribution of PSR 0540-69 to the LMC X--1 PDS flux, 
we first reanalyzed a \BS archival observation of 25 October 1996, from which 
Mineo et al. (1999) reported on the LECS/MECS properties of PSR 0540-69. 
In this pointing of PSR 0540-69, LMC X--1 is $25^{\prime}$ offset.
The 15--100 keV PDS count rate in our observation is $0.43\pm$0.07 cts/s,
while that in the 1996 observation is $0.34\pm$0.06 cts/s. 
The ratio between the two is $0.8\pm 0.2$, consistent with the ratio of the
effective areas at the position of LMC X--1 in the two cases, that is 0.68. 
This suggests that most of the PDS signal comes from LMC X--1. 

An alternative way to compare the amount of flux in the PDS from
each source, is to extrapolate to the PDS range the model spectra used in the 
the MECS analysis of our data.
We fitted the MECS data of PSR 0540-69 to an absorbed PL
with $N_{\rm H}$ fixed to $4 \times 10^{21}$ cm$^{-2}$ (Finley et al. 1993). 
The best fit photon index is $\Gamma = 1.95^{+0.09}_{-0.08}$, 
consistent with previous determinations (Mineo et al. 1999, Finley et al. 1993).
We then extrapolated such model spectrum in the PDS, 
correcting the MECS counts by a factor $0.85\times 0.9\times 0.68 = 
0.52$, where the first term takes into account the different flux normalizations 
between the two intruments, the second term is due to the PSA correction 
applied during the data reduction, the third term is simply the collimator 
relative transmission. 
After this, we obtained a simulated count rate, in the range [15--60] keV, 
of $0.12\pm0.04$ cts/s,  
while the measured one (in the same energy range) is $0.25\pm0.03$ cts/s. 
This fact suggests that a fraction between 30\% and 80\% 
of the flux detected in PDS could be attributed to the pulsar emission. 

The same procedure was applied to the LMC X--1 data. 
We used a model with the flattest powerlaw slope ($\Gamma = 3.07$) allowed by the 
99\% confidence contours, in order to maximize its possible contribution in the PDS 
(corresponding to $N_{\rm H} = 0.73\times10^{21}$ cm$^{-2}$). 
The simulated 15--60 keV countrate is $(5.92\pm2.76)\times10^{-2}$ 
cts/s. Such a value could explain the remaining fraction of PDS counts, 
once that the possible pulsar contribution, as computed above is subtracted. 
These results suggest that the contribution to the hard X--ray flux from LMC X-- 1 
should be equal to or lower than the contribution from the nearby source PSR 0540-69.
This conclusion is consistent with our previous argument, based on direct  
measurement of the count rates with LMC X--1 on and off axis, 
only if we assume that LMC X--1 and PSR 0540-69 contribute equally to the PDS 
signal.

\clearpage

\begin{table*}
\begin{tabular}{cccccccc}
\multicolumn{8}{l}{{TABLE 1: Observation Log}}\\
&&&&&&&\\ 
\hline
\hline
&&&&&&&\\
\multicolumn{1}{c}{Source Name}
&\multicolumn{3}{c}{Exposure Time (ks)}&
&\multicolumn{3}{c}{Count rate (cts/s)}\\
&&&&&&&\\
& 
\multicolumn{1}{c} {LECS} & 
\multicolumn{1}{c} {MECS} &
\multicolumn{1}{c} {PDS$^{(a)}$} & &
\multicolumn{1}{c} {LECS$^{(b)}$} &
\multicolumn{1}{c} {MECS$^{(c)}$} &
\multicolumn{1}{c} {PDS$^{(d)}$}\\
&&&&&&&\\
\hline
&&&&&&&\\
{LMC X--1} & {14.4} & {38.2} & {21+22}& & 
{3.26$\pm{0.02}$} & {5.12$\pm{0.02}$} & {0.22$\pm{0.03}$}\\
&&&&&&&\\
{LMC X--3} & {17.4} & {41.1} & {19+19}& &
{3.37$\pm{0.01}$} & {4.44$\pm{0.01}$} & $<$11.25 (3$\sigma$)\\
&&&&&&&\\
\hline
&&&&&&&\\
\multicolumn{8}{l}{$^{(a)}$: exposure time is for the two collimators.}\\
\multicolumn{8}{l}{$^{(b)}$: 0.2--4 keV.}\\
\multicolumn{8}{l}{$^{(c)}$: 1.8--10 keV (units 2+3).}\\ 
\multicolumn{8}{l}{$^{(d)}$: 15--60 keV (PSA correction included).}\\
\end{tabular}
\end{table*}

\begin{table}
\begin{tabular}{lcccc}
\multicolumn{5}{l}{{TABLE 2: Spectral Fits (absorbed disk blackbody)}}\\
&&&&\\
\hline
\hline
&&&&\\
\multicolumn{1}{l}{Source Name}
&\multicolumn{1}{c}{$N_{\rm H}$}
&\multicolumn{1}{c}{$kT_{\rm DBB}^{(a)}$}
&\multicolumn{1}{c}{$R_{\rm DBB}^{(b)}$}
&\multicolumn{1}{c}{$\chi^2/dof$}\\
&{($10^{22}$ cm$^{-2}$)} & {(keV)} & {(km)}
&\\
&&&&\\
\hline
&&&&\\
{LMC X--3} &
{$0.040^{+0.003}_{-0.001}$} &
{$1.04^{+0.01}_{-0.01}$} &
{$26.4^{+0.3}_{-0.2}$} &
{$283/216$}\\
&&&&\\
{LMC X--1 (total)} &
{$0.46$} &
{$0.97$} &
{$35.2$} &
{$708/216$}\\
&&&&\\
\hline
&&&&\\
\multicolumn{5}{l}{$^{(a)}$: temperature of the inner accretion disk radius 
$R_{\rm in}$.}\\ 
\multicolumn{5}{l}{$^{(b)}$: $R_{\rm DBB}=R_{\rm in}\,\sqrt{\cos{\theta}}$ 
for a distance of 55 kpc} \\
\end{tabular}
\end{table}

\begin{table}
\begin{tabular}{lcccccc}
\multicolumn{7}{l}{{TABLE 3: Spectral Fits (absorbed disk blackbody + power
law)}}\\
&&&&&&\\
\hline
\hline
&&&&&&\\
\multicolumn{1}{l}{Source Name}
&\multicolumn{1}{c}{$N_{\rm H}$}
&\multicolumn{1}{c}{$kT_{\rm DBB}^{(a)}$}
&\multicolumn{1}{c}{$R_{\rm DBB}^{(b)}$}
&\multicolumn{1}{c}{$\Gamma$}
&\multicolumn{1}{c}{$K_{\rm PL}^{(c)}$}
&\multicolumn{1}{c}{$\chi^2/dof$}\\
&{($10^{22}$ cm$^{-2}$)} & {(keV)} & {(km)}
&&&\\
&&&&&&\\
\hline
&&&&&&\\
{LMC X--3} &
{$0.07^{+0.01}_{-0.01}$} &
{$1.03^{+0.01}_{-0.01}$} &
{$26.5^{+0.4}_{-0.3}$} &
{$2.73^{+0.17}_{-0.20}$} &
{$0.015^{+0.004}_{-0.005}$} &
{$238/214$}\\
&&&&&&\\
{LMC X--1 (total)} &
{$0.81^{+0.03}_{-0.05}$} &
{$0.92^{+0.01}_{-0.01}$} &
{$35.8^{+1.3}_{-1.1}$} &
{$3.26^{+0.11}_{-0.12}$} &
{$0.22^{+0.05}_{-0.05}$} &
{$212/193$}\\
&&&&&&\\
{LMC X--1 (Part 1)} &
{$0.85^{+0.08}_{-0.08}$} &
{$0.89^{+0.01}_{-0.01}$} &
{$40.4^{+1.5}_{-1.4}$} &
{$3.46^{+0.17}_{-0.20}$} &
{$0.24^{+0.08}_{-0.07}$} &
{$208/195$}\\
&&&&&&\\
{LMC X--1 (Part 2)} &
{$0.77^{+0.06}_{-0.07}$} &
{$0.97^{+0.02}_{-0.02}$} &
{$28.8^{+2.0}_{-1.7}$} &
{$3.01^{+0.14}_{-0.16}$} &
{$0.20^{+0.06}_{-0.05}$} &
{$261/195$}\\
&&&&&&\\
\hline
&&&&&&\\
\multicolumn{7}{l}{$^{(a)}$: temperature of the inner accretion disk radius 
$R_{\rm in}$.}\\ 
\multicolumn{7}{l}{$^{(b)}$: $R_{\rm DBB}=R_{\rm in}\,\sqrt{\cos{\theta}}$ 
for a distance of 55 kpc} \\
\multicolumn{7}{l}{$^{(c)}$: photons/cm$^2$/s/keV at 1 keV.}\\
\end{tabular}
\end{table}

\clearpage

\end{document}